\documentclass[conference]{IEEEtran}
\usepackage[utf8]{inputenc}
\usepackage{amsmath,amssymb,amsfonts}
\usepackage{graphicx}
\usepackage{textcomp}
\usepackage{xcolor}
\usepackage{tabto}
\usepackage{dirtytalk}
\usepackage{graphicx}
\usepackage{setspace}
\usepackage{mathtools}
\usepackage{accents}
\usepackage{enumitem}
\usepackage{hyperref}
\usepackage{longtable}
\usepackage{fixmath}
\usepackage{wrapfig}
\usepackage{color}
\usepackage{algorithm,algpseudocode}
\usepackage{comment}
\usepackage{subfig}
\usepackage{soul}
\usepackage{tabularx}
\usepackage{multirow}
\usepackage{lipsum}
\usepackage{graphicx}
\usepackage{siunitx}
\usepackage{amssymb}
\usepackage[bottom]{footmisc}

\usepackage{array}
\def\BibTeX{{\rm B\kern-.05em{\sc i\kern-.025em b}\kern-.08em
    T\kern-.1667em\lower.7ex\hbox{E}\kern-.125emX}}
    
\IEEEoverridecommandlockouts    

\begin{document}

\title{Deep Reinforcement Learning for Network Slicing with Heterogeneous Resource Requirements and Time Varying Traffic Dynamics\thanks{A shorter version of the paper will appear in CNSM 2019.}}

\author{\IEEEauthorblockN{Jaehoon Koo}
\IEEEauthorblockA{\textit{Northwestern University} \\
Evanston, IL, USA \\
jaehoonkoo2018@u.northwestern.edu} \vspace{.5cm}

\IEEEauthorblockN{Muntasir Raihan Rahman}
\IEEEauthorblockA{\textit{Nokia Bell Labs} \\
Murray Hill, NJ, USA  \\
muntasir.rahman@nokia-bell-labs.com}

\and 

\IEEEauthorblockN{Veena B. Mendiratta}
\IEEEauthorblockA{\textit{Nokia Bell Labs} \\
Naperville, IL, USA \\
veena.mendiratta@nokia-bell-labs.com} \vspace{.5cm}

\IEEEauthorblockN{Anwar Walid}
\IEEEauthorblockA{\textit{Nokia Bell Labs} \\
Murray Hill, NJ, USA  \\
anwar.walid@nokia-bell-labs.com} 
}

\maketitle

\begin{abstract}
 
Efficient network slicing is vital to deal with the highly variable and dynamic characteristics of network traffic generated by a varied range of applications.
The problem is made more challenging with the advent of new technologies such as 5G and new architectures such as SDN and NFV. 
Network slicing addresses a challenging dynamic network resource allocation problem where a single network infrastructure is divided into (virtual) multiple slices
to meet the demands of different users with varying requirements, 
the main challenges being --- the traffic arrival characteristics and the job resource
requirements (e.g., compute, memory and bandwidth resources)
for each slice can be highly dynamic. 
Traditional model-based optimization or queueing theoretic modeling becomes intractable with the high reliability, and stringent bandwidth and latency requirements imposed by 5G technologies. In addition these approaches lack adaptivity in dynamic environments.
We propose a deep reinforcement learning approach to address this dynamic coupled resource allocation problem. 
Model evaluation using both synthetic simulation data and real workload driven traces demonstrates that our deep 
reinforcement learning solution improves overall resource utilization, 
latency performance, and demands satisfied as compared to a baseline equal-slicing strategy.
\end{abstract}


\section{Introduction}
\label{section:introduction}

The challenges introduced by new softwarized technologies such as network function virtualization (NFV)~\cite{mijumbi2016network} and 
software-defined networking (SDN)~\cite{mckeown2009software}, and 
new network architectures such as 5G,
are driving network transformations that radically change the way operators manage
their networks and orchestrate network services.
The network architectures come with a diverse range of capabilities
and requirements, including 
massive capacity (e.g., Massive MIMO~\cite{larsson2017massive}),
ultra low latency (ULL), ultra high reliability, and 
support for massive machine to machine communications (M2M) in the context of Industry 4.0~\cite{wiki:industry4.0}.
Networks are being transformed into programmable, software-driven, service-based and holistically managed systems, accelerated via enablers such as NFV, SDN, and 
mobile edge computing. 
In order to consolidate multiple networks with varied requirements, the 5G architecture must exploit network virtualization and programmability. 
This introduces the problem of network slicing~\cite{5GSM2017}, where a single network infrastructure is divided into multiple sub-networks, and each slice can be operated by different parties.

Network slicing can be modeled as a dynamic resource allocation problem.
Once the network is sliced into multiple sub-networks with SLA requirements on latency, bandwidth, reliability, etc., the underlying infrastructure operator needs to ensure
that the SLAs for each slice are guaranteed. 
These strict SLA guarantees need to be maintained despite variability in the slice  request arrivals and request resource requirement distributions. 
For example, if one of the slices is owned by a video streaming operator and requires low latency during a popular gaming event, then the slice resource allocation needs to be dynamically adjusted as a burst of requests arrive. 
While traditionally such network resource allocation problems have been solved using analytical queueing theoretic and optimization methods~\cite{Neely:2010:SNO:1941130, Shakkottai:2007:NOC:1345033.1345034, Srikant:2014:CNO:2636796},
given the complexity and scale of modern networks
such methods are not feasible. 
Thus we resort to approximate black-box models using machine learning, namely Reinforcement Learning (RL)~\cite{sutton2018reinforcement}
%Network slicing algorithms based on RL can generalize to new and unknown network traffic dynamics, unlike static algorithms that must be tuned for new network environments and traffic patterns.

Reinforcement learning is a computational approach for
goal-directed learning and decision making,
with the goal being to select actions to maximize future rewards \cite{sutton2018reinforcement}.
The emphasis is on learning by an agent
(with the capacity to act) where each action influences the agent's future state,
through direct interaction with its environment, without the need for exemplary supervision or complete models of the environment.
RL methods are self-customizable, as they can adapt the mapping from state to actions to maximize expected rewards in response to changing environmental conditions,
for example, network environment (wireless vs wired), traffic dynamics, job size distribution, etc.
Recent developments in RL employ deep neural networks to learn complex patterns in the experienced state, and can select high reward actions (e.g., AlphaGo~\cite{silver2017mastering}). 
Our hypothesis is that deep RL can be used to learn good network slicing strategies 
by learning over simulated and trace driven workload data, and 
such learned policies can be applied for real network slicing deployments. 

To satisfy a network slice resource request, a network operator needs to simultaneously allocate heterogeneous resource requests such as compute, bandwidth and storage, 
adding additional complexity to the problem.
To summarize, we face three main challenges for efficient dynamic resource allocation for network slicing.

\begin{enumerate}
    \item Unknown request arrival process and request resource requirements.
    \item Heterogeneous resource requirements for each slice (e.g., CPU, bandwidth, memory, etc.).
    \item Finite resource capacities.
\end{enumerate}

While existing solutions deal with each challenge separately, 
we propose a unified solution using deep RL that can simultaneously deal with varying and unknown traffic arrival dynamics, heterogeneous resource requirements, and finite resource capacities.  
In our formulation, each network slice has both 
bandwidth and compute resource allocation requirements, 
where the distribution of request arrivals and
the amount of resources requested is not known apriori.
Our main contributions are as follows:
\begin{enumerate}[label=\roman*)]
 \item Mathematical formulation of the network slicing resource allocation  RL problem as a Markov Decision Process (MDP) where the 
 constrained multi-resource optimization problem is formulated for service upon arrival and batch service.
 \item A policy-gradient method to solve this problem based on the popular REINFORCE~\cite{silver2017mastering} algorithm where we use a deep neural network architecture as the function approximator for learning the optimal policy. 
 \item Experimental study with varying resource budgets
 using both simulated and real data-sets
 to evaluate our proposed solution versus an equal slicing strategy.
\end{enumerate}

Our RL framework for network slicing creates new opportunities for network and infrastructure operators, as they can train the network slicing models based on their target network dynamics, network heterogeneity, and SLA requirements. 
The models can be trained off-line using simulated and workload data, and then deployed in real-time for slicing resource allocation decisions.

%{\color{red} Muntasir: If we can add some comments as to why we need non-linear models as opposed to simpler linear models that would be good}

%{\color{red} Jaehoon: (Also it would be good to discuss why stateless bandits is not sufficient for this problem) We have RL problems as a simple RL problem, called a contextual bandit problem, where state and action influence reward, however, state does not necessarily depend on action; and a full RL problem, where state depends on action \cite{Agarwal2016AMT}. In this study, we formulate the full RL problem since we assume to have buffers. In our formulation, state, buffer level, depends on action, resource allocation, and both action and state affect reward, QoS and resource use costs.
%}

The remainder of the paper is organized as follows. 
Section~\ref{section:network-slicing} gives an overview of the network slicing architecture in 5G. 
Section~\ref{section:rl} gives an introduction to RL, 
including markov decision process (MDP) models and policy gradient algorithms. 
In Section~\ref{section:models} we formulate the models for 
dynamic resource allocation for network slicing as an MDP, and give details of the policy gradient learning algorithms. 
Section~\ref{section:data} gives the details of the workload traces for CPU and bandwidth that are used in the experiments. 
Section~\ref{section:perf-eval} presents the experimental results of our proposed models using both simulated and real data-sets. 
Finally we discuss related work in Section~\ref{section:related-work} and 
conclude in Section~\ref{section:conclusions}.

\section{Network Slicing}
\label{section:network-slicing}
Network slicing in 5G is an emerging technology area and, hence, presents many opportunities
as well research challenges. 
In \cite{7Li2017} the authors propose a framework for the technology, and discuss challenges 
and future research directions listing the efficient allocation of slice resources as a
challenging problem to be addressed algorithmically.

The basic concept of network slicing is a virtual network architecture
running multiple
logical networks on a common shared physical infrastructure.
Each network slice represents an independent virtualized end-to-end network
customized to meet the specific needs of an application, service, device, customer or operator \cite{5GSM2017}.
It comprises a set of logical (software) network functions that support the requirements of the particular use case.
where each function is optimized to provide the resources and network topology for the specific service and traffic that will use the slice. 

\begin{figure}[htbp]
\centerline{\includegraphics[scale=0.3]{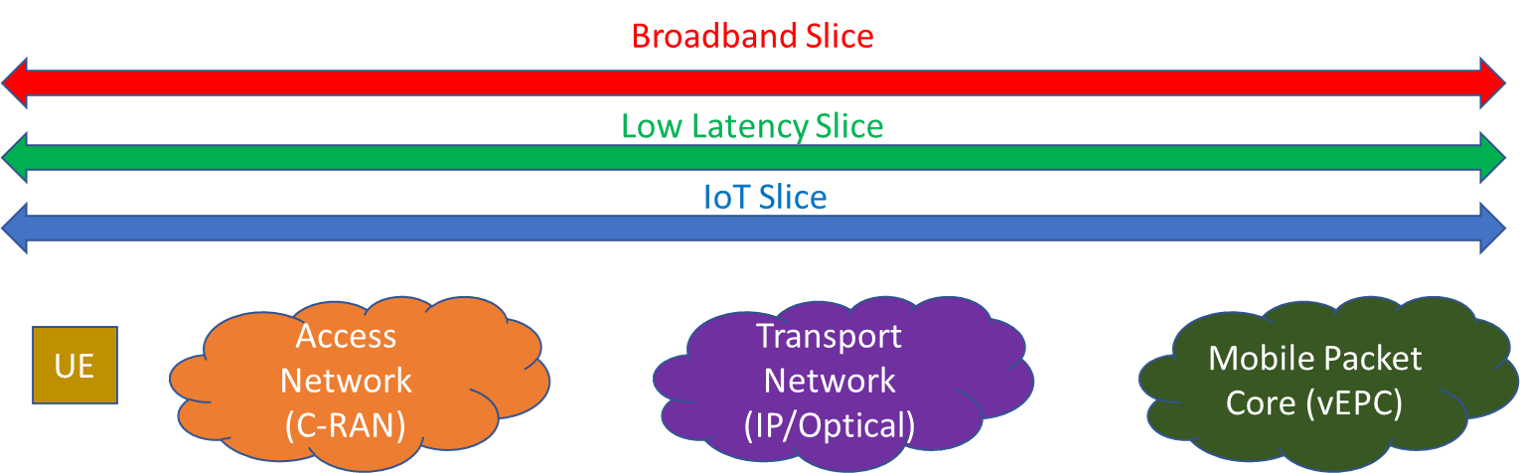}}
\caption{Example of 5G network slicing}
\label{fig-slice}
\end{figure}

A key benefit of network slicing is that it provides an end-to-end virtual 
network encompassing compute, bandwidth and storage functions. 
The objective is to allow a network operator to partition its network resources to allow for different types users or tenants to multiplex over a single physical infrastructure.
A typical example used in the 5G literature is the following:
run Internet of Things (IoT), Mobile Broadband (MBB), and very  vehicular communications applications on the same network. 
IoT will typically have a large number of devices each with low throughput, 
while MBB will have a smaller number of devices with high bandwidth content, and 
vehicular communications will have stringent requirements of low-latency.
The goal of network slicing is to enable partitioning of the physical network at an end-to-end level
so as to allow optimum grouping of traffic, tenant isolation, and configuration of resources at a macro level.
In Figure~\ref{fig-slice} we show an example of
5G network slices spanning the access, transport and mobile packet core network domains \cite{rost2017network}.

\section{Background on RL and Deep RL}
\label{section:rl}
In this section, we give a brief introduction of RL, specifically the use of neural networks
for RL function approximation, and the policy gradient learning algorithm.

\subsection{Basic Reinforcement Learning Model}
In RL, an agent interacts with an environment. The agent has a set of discrete or continuous actions to choose from, and the action can influence the next state of the environment.
We consider the full RL model, where actions can influence state.
This is unlike the bandit framework, where actions do not influence state, and,
at every instance, the agent only chooses the best action independent of how it may impact the state. 

\begin{figure}[htbp]
\centerline{\includegraphics[scale=0.4]{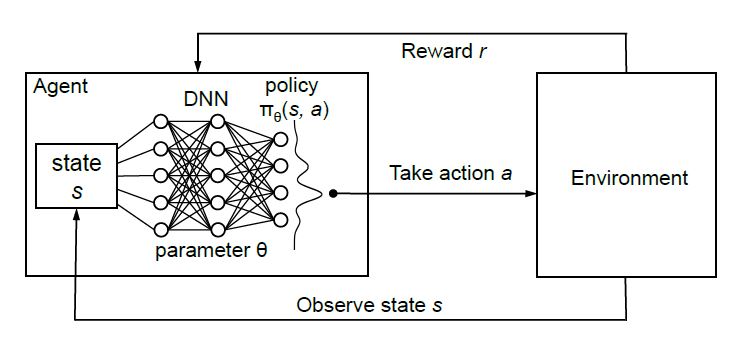}}
\caption{Deep Reinforcement Learning \cite{8Mao2016}}
\label{fig-rl}
\end{figure}

At each time step $t$, the RL agent observes a local copy of the environment's state $s_t$, and selects an action $a_t$. At the next time step $t+1$, the agent observes a reward $r_t$ which represents the cost/reward for taking the last action. The agent also observes the next state $s_{t+1}$. 
We make the markovian assumption that the future state ($s_{t+1}$) only depend on the current state ($s_t$), and 
also that the dynamics are stationary and do not change over time. 
The agent's goal is to maximize the expected cumulative discounted return:

\begin{equation}
\label{eq:exp-cum-reward}
    R_t=\mathbb{E}[\sum_{t}\gamma^t \cdot r_t], \text{ for } \gamma \in [0,1)
\end{equation}
where the parameter $\gamma$ is the discount factor and it weighs immediate reward with possibilities of better rewards in the future. 
It is also a mathematical trick to ensure cumulative rewards converge for infinite horizon problems. 
Figure~\ref{fig-rl} summarizes the full RL model.

\subsection{Policy}

The agent chooses actions based on a learned policy, which is a probability distribution over actions for any state, $\pi: \pi(s,a) \rightarrow [0,1]$, 
where $\pi(s,a)$ denotes the probability of taking action $a$ in state $s$, and $\sum_a\pi(s,a)=1, \forall s$. 
In most practical scenarios, especially network environments, there are exponentially many possible state,action (s,a) pairs 
(see section~\ref{section:models}). 
Thus modern scalable methods eschew tabular representations in favor of function approximators~\cite{sutton2018reinforcement}. 
A function approximator is parameterized by $\theta$, and
policies are denoted by $\pi_{\theta}(s,a)$. 
Typically policy approximations learn to cluster the behavior of similar states such that, given a new state,
the approximator can find the action for the closest state seen so far.

There is considerable research on various types of function approximators. 
Any popular supervised learning framework such as SVM can be used. 
Recently a popular choice is to use a deep neural network (DNN) as the function approximator~\cite{sutton2018reinforcement}
which we also use in our models.
An attractive feature of DNNs is that the model automatically learns the best representation of the feature space. 

\subsection{Policy Gradient Methods} We utilize a class of policy learning algorithms called policy gradient methods, that try to learn an optimal policy using gradient descent (or ascent)~\cite{sutton2018reinforcement}. The objective is to maximize the expected cumulative reward (Equation~\ref{eq:exp-cum-reward}); the gradient of this objective is given by:

\begin{equation}
    \nabla_{\theta} \mathbb{E}_{\pi_{\theta}} [\sum_{t=0}^{\infty} \gamma^t r_t] = \mathbb{E}_{\pi_{\theta}} [\nabla_{\theta} \log \pi_\theta (s, a) Q^{\pi_\theta}(s,a)],
\end{equation}
where $Q^{\pi_{\theta}}$ represents the expected cumulative discounted reward from selecting the action $a$ in state $s$ and then following policy $\pi_{\theta}$.
This class of methods estimate the gradient by sampling trajectories of policy executions and 
obtaining a reward estimate $v_t$ for the trajectory,
and subsequently update the policy parameters using gradient ascent using the following equation:

\begin{equation}
    \theta \leftarrow \theta + \alpha \sum_{t} \nabla_{\theta} \log \pi_{\theta} (s_t, a_t) v_t,
\end{equation}
where $\alpha$ is the gradient ascent step size. This results in the REINFORCE algorithm~\cite{silver2017mastering}, which we used for learning in our system. 
The pseudo-code of our implemented training algorithm is in Section~\ref{section:models}.

%\begin{algorithm} 
%\caption{REINFORCE} %\cite{Silver2015}
%    \label{alg:1}
%    \begin{algorithmic}[1]
%    \State {Initialize $\theta$ arbitrarily}
%        \For{each episode $\{s_1, a_1, r_2, \ldots, s_{T-1}, a_{T-1}, r_{T-1}\} \sim \pi_{\theta}$}
%            \For {$t=1$ to $T-1$}
%            \State {$\theta \leftarrow \theta + \alpha \sum_t \nabla_{\theta} \text{log} \pi_{\theta} v_t $}
%            \EndFor
%        \EndFor
%    \end{algorithmic}
%\end{algorithm}

%{\color{red}MUNTASIR: I need to see the implemented policy gradient code to write some details here}

An advantage of policy-gradient methods is that they directly search the policy space and can be quite generic
so that a general algorithm can be adapted to various scenarios without 
significant modification. 
Policy gradient techniques have found success in robotics, game playing, and cloud resource allocation domains. 
There are other popular learning algorithms such as Q-learning which can be used. Below we briefly describe the differences among policy gradient algorithms such as REINFORCE and value iteration methods such as Q-learning and indicate why we choose policy gradient for our system.

First, both classes of algorithms can solve general MDPs and can converge to optimal policies. 
However, their internal structures are different.
The fundamental difference is in the approach to action selection, both whilst learning and as the output (the learned policy). 
In Q-learning, the goal is to learn a single deterministic action from a discrete set of actions by finding the maximum value. 
With policy gradients, and other direct policy searches, the goal is to learn a map from state to action, which can be stochastic, and can work in continuous action spaces.
As a result, policy gradient methods can solve problems that value-based methods cannot, especially in scenarios with large and continuous action spaces, and stochastic policies.
In our RL formulation (details in Section~\ref{section:models}), 
we deal with real valued resource allocations
with large action spaces, for example, compute and bandwidth.
Due to these benefits we selected policy gradient methods in our RL agents, 
though our system is general enough to use alternative RL algorithms as well.

%\begin{enumerate}
 %   \item Large and continuous action space
  %  \item Stochastic policies. A value-based method cannot solve an environment where the optimal policy is stochastic requiring specific probabilities. That is because there are no trainable parameters in Q-learning that control probabilities of action, the problem formulation in temporal difference learning assumes that a deterministic agent can be optimal.
%\end{enumerate}

\section{Proposed Models}
\label{section:models}
In this section we present models for allocating bandwidth and Virtual Machines (VM) to network slices. We formulate the resource allocation problem, and describe this in RL settings. We present two models for different service types; service upon arrival and batch service. 

\subsection{Service upon arrival}

\begin{figure}[htbp]
\centerline{\includegraphics[scale=0.42]{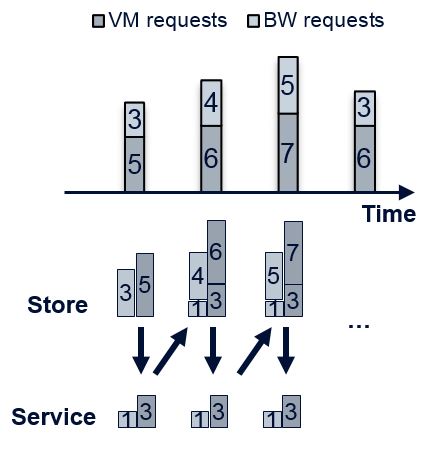}}
\caption{Service upon arrival}
\label{fig-SA}
\end{figure}

For service upon arrival,
consider a mobile network that receives resource requirements for bandwidth and VMs from a set $K$ of classes during the time horizon $T$. It is assumed that requests for bandwidth and VMs have the same arrival process for each class. However, the quantities of the resource requests have different and independent distributions. We assume that we have infinite buffers for both resources to hold received requests. The controller determines  resource allocations for each class (slice) for the whole network at any arrival of each set of requests. 
Fig. \ref{fig-SA} presents the request arrival and service processes for
the service upon arrival formulation.

A set of bandwidth and VM requests for each class has a arrival process $\rho_i (t), i\in K, t\in T$. 
Each arrival has different amounts of arriving requests for bandwidth and VMs from each class. 
For bandwidth we have amounts of the requests, $x_i (t) \sim M_i (t) | \rho_i (t)$ and buffer levels $\beta_i(t), i \in K, t \in T$.\footnote{{\color{black}The buffer level can be computed by $\beta_i(t) = x_i (t) + \beta_i(t-1) - b_i (t-1)$.}} 
Similarly, for VMs, we have amounts of the requests, $y_i (t) \sim N_i (t) | \rho_i (t)$ and buffer levels $\delta_i (t), i \in K, t \in T$. 
Furthermore, we have resource allocations by a controller, $b_i (t) | x(t),\beta (t)$ and $v_i(t) | y(t),\delta(t), i \in K, t \in T$ for bandwidth and computing resources.

Our goal is to maximize the Quality of Service (QoS) with respect to bandwidth and VM requests and minimize resource usage costs. Measuring QoS as delays to process received requests, we achieve our goal by solving the following optimization problem:
%{\color{blue}
\begin{equation}
\label{model:eq1}
\begingroup\small
\begin{aligned}
& \underset{}{\text{minimize}}
& & \mathbb{E} \big[ \sum_{t=0}^{T} \gamma^{t} \mathcal{L}(\mathbf{b}(t),\mathbf{x}(t),\boldsymbol{\beta}(t),\mathbf{v}(t),\mathbf{y}(t),\boldsymbol{\delta}(t))\big] \\ % |\mathbf{x}(t),\mathbf{y}(t)
\end{aligned}
\endgroup
\end{equation}%}
where $\mathcal{L} = L_{\text{QoS}}(\boldsymbol{\beta},\boldsymbol{\delta}) + w L_{\text{Res}}(\mathbf{b},\mathbf{v})$, and $L_{\text{QoS}}$ integrates delays in processing bandwidth and VM requests measuring buffer levels, $L_{\text{Res}}$ is a cost for bandwidth and computing resources, and $\gamma$ and $w$ are a discount and a balance factors. 

We solve the above optimization problem with RL algorithms. TABLE \ref{table-rl} shows the state, action and reward for the RL formulation. 

\begin{figure}
    \begin{algorithmic}[1]
        \State {Initialize $\theta_{b}, \theta_{v}$}
        \For{$\text{episode}=1$ to all episodes}
            \For {$t=1$ to $T-1$}
            \If{$\sum_{i=1}^{|K|} b_{i}^{t} \leq B^{t}$}
            \State{$b_{i}^{t} \leftarrow \frac{b_{i}^{t}}{\sum_{i=1}^{|K|} b_{i}^{t}} B^{t}$}
            \EndIf
            \State {$\theta_b \leftarrow \theta_b - \alpha \nabla_{\theta_b} J_t $}
            \If{$\sum_{i=1}^{|K|} v_{i}^{t} \leq C^{t}$}
            \State{$v_{i}^{t} \leftarrow \frac{v_{i}^{t}}{\sum_{i=1}^{|K|} v_{i}^{t}} C^{t}$}
            \EndIf
            \State {$\theta_v \leftarrow \theta_v - \alpha \nabla_{\theta_v} J_t $}
            \EndFor
        \EndFor
    \end{algorithmic}
    \caption{REINFORCE~\cite{Williams:1992:SSG:139611.139614} Algorithm for Network Slicing}
    \label{fig:alg-rl}
\end{figure}

Deep neural networks are introduced as a policy agent for determining the policy $\pi(s,a)$. 
We assign a separate agent for each of the resources, namely CPU and bandwidth.
The  neural network agents feed the input feature vector: $<R,B,A>$, where $R$ denotes the amount of received resource requests, $B$ denotes the buffer level, and $A$ denotes the last request arrival time, and computes as output the slice resource allocation. %, {\color{blue} JK:  $\pi_{\theta}(<R, B, A>,a)$.} %, $S = f(<R, B, A>)$. 
We solve our RL models by applying a class of RL algorithms, policy gradient methods, that learn by performing gradient-descent on the policy parameters \cite{sutton2018reinforcement}. The algorithm based on REINFORCE~\cite{Williams:1992:SSG:139611.139614} is shown in Figure~\ref{fig:alg-rl}. In this algorithm, $J$ denotes the objective function, which is the expected cumulative discounted loss as shown in Eq.~\ref{model:eq1}, and $\theta_{b}, \theta_{v}$ are learning parameters of policy agents for each resource.
Each neural agent consists of multiple hidden layers and one output layer. 
The leaky ReLU is adopted as an activation function at hidden and output layers,
\[
    h(x)= 
\begin{cases}
    x,& \text{if } x > 0\\
    ax,              & \text{otherwise}
\end{cases}
\]
with a small constant $a$. 
The leaky ReLU is selected as it produces positive allocations, and
resolves a difficulty of ReLU when units are not active by allowing a small, positive gradient \cite{Maas2013}. 

\begin{table} [H]
\caption{State, action and reward}
    \begin{tabularx}{\columnwidth}{ | l | X | X | }
    \hline
    & Service upon arrival & Batch service \\ \hline
    State  & Resource requests arrivals, and buffer levels since last arrival time& Resource requests arrivals, and buffer levels since last service time\\ \hline
    Action & Resource allocation to each slice & Resource allocation to each slice  \\ \hline
    Reward & -(Delays in processing requests and resource use costs) & -(Delays in processing requests loss and resource use costs) \\
    \hline
    \end{tabularx}
\label{table-rl}
\end{table}

\subsection{Batch service}
\begin{figure}[htbp]
\centerline{\includegraphics[scale=0.42]{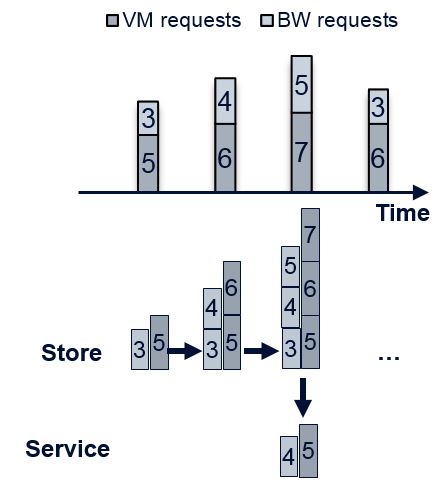}}
\caption{Batch service}
\label{fig-BS}
\end{figure}

Next we formulate the resource allocation problem for a batch service strategy, where the slice orchestrator wakes up periodically and processes all the requests in the queue. 
Selecting between service upon arrival and batch service is a trade-off between increased delay (waiting time in queue) and reduced scheduling overhead (allocate resources at scheduled times). 
In batch service, received resource requests are held in buffers, and 
resources are allocated for processing at pre-determined times. 
The same arrival process and assumptions are applied to this formulation as for the
service upon arrival formulation. 
Fig. \ref{fig-BS} presents an example request arrival and service process  for the batch mode of operation. 

The same optimization problem is solved for the batch formulation as for the service upon arrival formulation. The MDP formulation for this model is shown in Table \ref{table-rl}, however
a different neural network structure is proposed for a policy agent in this batch mode. For this mode, each deep neural network agent per class uses statistical measures (mean, max and standard deviation) of $R$ (the amount of received resource requests), $B$ (the buffer level), and $A$ (the request arrival time) computed from the requests in a single batch. 
%Thus in this case, output resource allocation is 
%\begin{equation}
 %S=f(<\mu(R), \max(R), \sigma(R), \mu(B), \max(B), \sigma(B), \mu(A), \max(A), \sigma(A)>),    
%\end{equation}
%where $\mu(.)$ and $\sigma(.)$ denote mean and standard deviation respectively . 
This allows the batch service to capture the dependencies among the requests in each batch, and prevents the allocation to simply sum up per request allocations.
%While assigning an agent for a class by resource, the neural network agents input statistics (mean, max, and standard deviation) of received resource requests, buffer levels, and scheduled service times, and
%output resource allocation for each slice. 
%This structure is constructed to prevent the batch service model from simply being a summation of service upon arrival. 
The hidden and output layers for the neural network agents in this batch model are similar to the service upon arrival model. 

%\subsection{Budget constraints}

We compare our proposed algorithms to an equal slicing policy. This baseline strategy fairly divides the resources among each slice. Such a policy only makes sense when there is a finite budget for resources. Thus in our RL algorithms, we also impose budget constraints for each resource type to make a fair comparison, however our proposed algorithms are generic and can also work when there is no bound on resource capacity.

%We present a baseline model in this section. 
 %To this end, we should have budgets for each resource. 
 
 We introduce budget constraints to Equation (\ref{model:eq1})
\begin{equation*}
\begin{aligned}
\sum_{i=1}^{|K|} b_{i} \leq B, \text{ and } \sum_{i=1}^{|K|} v_{i} \leq C \\
\end{aligned}
\end{equation*}
where $B$ is the budget for bandwidth and $C$ is the budget for computing resources.
Throughout the whole time horizon, we do not allow resource allocations larger than the budgets.
To implement budget constraints, we project obtained allocations from neural agents proportionally when the sum of the allocations exceeds the budgets,
\begin{equation*}
\begin{aligned}
b_{i} \leftarrow \frac{b_{i}}{\sum_{i=1}^{|K|} b_{i}} B \text{ and } v_{i} \leftarrow \frac{v_{i}}{\sum_{i=1}^{|K|} v_{i}} C. \\
\end{aligned}
\end{equation*}

The budget constraints are incorporated in Algorithm~\ref{fig:alg-rl}.

\section{Data for Models}
\label{section:data}
In this section we briefly discuss the workload traces for CPU and bandwidth
requests that are used for the evaluation of the algorithms
(Section~\ref{section:perf-eval}).

\textbf{Facebook Workload Trace.}
SWIM~\cite{swim} is a workload generator for map-reduce cluster traces. 
The SWIM repository contains real workload traces from production clusters. We utilize the SWIM workload suite for Facebook map-reduce clusters. The traces contain both job arrival times and job sizes (map input bytes). 
The arrival data shows that the arrival process is generally bursty and is more realistic than using the Poisson arrival process for simulating the job arrival process. Each workload is for 1 hour of production cluster usage.
In our experiment we use 3 different traces for each slice. 
For each slice we run through the trace by simulating an event of job submission at corresponding arrival time from the trace, and 
use the job size at that time instance. 

\textbf{Bandwidth Data Trace.}
We use a 4G LTE trace of bandwidth data from~\cite{Hooft2016} for the experiments. Since we have a continuous bandwidth trace, we only sampled the bandwidth at specific instances.

\begin{figure} 
    \centering
  \subfloat[]{%
       \includegraphics[width=0.3\linewidth]{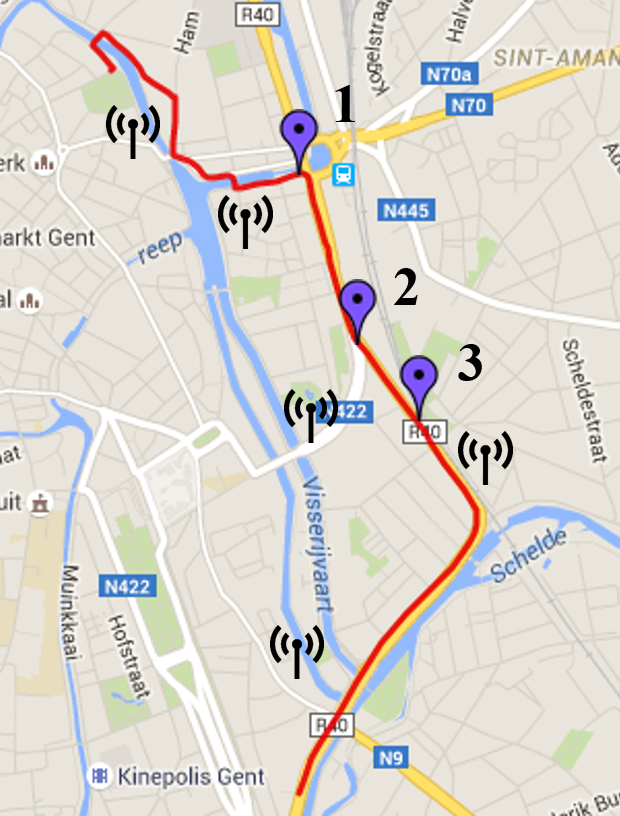}}
    \hfill
  \subfloat[]{%
        \includegraphics[width=0.65\linewidth]{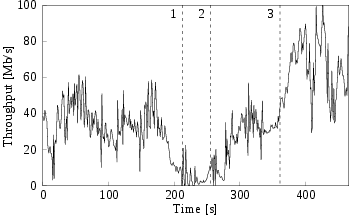}}
  \caption{Bandwidth Traces \cite{Hooft2016}}
  \label{sim:fig} 
\end{figure}

\textbf{Merging CPU and Bandwidth Traces.}
Since the traces for CPU and bandwidth were separate, 
we had to make some assumptions to combine the two traces for our experiments.
For each arrival time in the Facebook trace, 
we take the maximum bandwidth value in the interval 
$[current~job~arrival~time~-~previous~job ~arrival~ time]$.

\section{Performance Evaluation}
\label{section:perf-eval}
In this section we present the scenarios for experimentation, and 
the  results obtained with simulated and real data.
Model implementations for all the experiments are done in Python TensorFlow using
the GeForce GTX 1080 and Intel(R) Core(TM) i7-6850K CPU @ 3.60GHz. 

\subsection{Scenarios}
For analysis, we create scenarios with 4 levels of the resource budget:
smaller, small, large, and larger.
Each arriving request requires two resources:
bandwidth and compute (expressed as number of VMs). 
The budget size is determined using the mean and standard deviations of the resource request distributions. For service upon arrival, 
\begin{equation*}
\begin{aligned}
 B = \sum_{i=1}^{|K|} \mu_i + c \: \sigma_i, \text{ and } C = \sum_{i=1}^{|K|} \nu_i + c \: \eta_i \\
\end{aligned}
\end{equation*}
where $(\mu_i, \sigma_i)$ and $(\nu_i, \eta_i)$ for class $i$ are the
mean and standard deviation respectively of request distributions for the bandwidth and VMs.
$c$ as 0, 1, 2, and 3 for scenarios of smaller, small, large, and larger resource budgets respectively. 
Batch service requires a larger budget since arriving requests wait in the buffer till the next batch service time. 
To this end, we multiply the service time interval and 
the request arrival rate to the budget of the service upon arrival 
such that
$B_{\text{batch service}} = \sum_{i=1}^{|K|} \tau_i \: \lambda_i \: (\mu_i + c \: \sigma_i)$ where, for class i,
$\tau_i$ is the service time and $\lambda_i$ 
is the arrival rate.

\subsection{Simulated Data}

The models are first validated by conducting experiments on the simulated data.
The data is generated in the same format as the real data described in Section \ref{section:data}. 
There are three classes of requests each with different characteristics of bandwidth and VM requests.
Request arrivals are assumed to follow a Poisson distribution with the average number of events per interval~=~2, and 
the resource amounts of the requests follow uniform distributions as shown in 
Table \ref{table:sim-amount}. 

\begin{table} [ht]
\centering
\caption{Simulated data: Request amounts $\mathcal{U}$(a,b)}
    \begin{tabular}{lcc|cc}
    \hline
          & \multicolumn{2}{c|}{Bandwidth}  & \multicolumn{2}{c}{VM}\\ \cline{2-5}
          & a & b & a & b \\
    \cline{1-5}
    Class 1  & 100  & 150 & 500  & 600 \\
    Class 2  & 100  & 200 & 1000 & 1500 \\ 
    Class 3  & 300  & 500 & 1000 & 2000 \\     \hline
  \end{tabular}
\label{table:sim-amount}
\end{table}

The bounds for the distributions are based on the assumption that each class has different characteristics in terms of the mean and standard deviation, 
e.g. class 2 has a larger standard deviation than class 1 for bandwidth. 
For the hyper-parameters, different settings were tested:
learning rates from 0.1 to 0.001, the number of layers from 1 to 3, and 500 and 1000 units per layer.
Based on the results the following settings were chosen.
For service upon arrival, we generate 1,000 episodes for training, and 100 episodes for test. 
For batch service, we generate 5,000 episodes for training, and 100 episodes for test. 
The service time interval is set at 10 for batch service, i.e. it allocates
resources every 10 time units. 
We put the same weights on delays and resource use costs, $w=1$.
In both cases, each neural agent has 3 hidden layers with 1,000 units. 
The Adam optimizer \cite{kingma2014adam}, an 
algorithm for first-order gradient-based optimization of
stochastic  objective  functions  based  on  adaptive  estimates  of  
lower-order  moments, is used for the gradient optimization using a 
learning rate of 0.001. 
\begin{comment}
The method is straightforward to implement, is computationally
efficient,
has low memory requirements, is invariant to diagonal re-scaling of the gradients,
and is well suited for problems that are large in terms of data and/or parameters.
The method is also appropriate for non-stationary objectives and
problems with very noisy and/or sparse gradients.
\end{comment}

Using the simulated data, 
Table \ref{table:sim-loss} shows the results, of the proposed 
models (\textbf{NN}) compared with the equal slicing strategy (\textbf{ES}),
in terms of the expected rewards for each class and resource. 
with the winners shown in \textbf{bold}. 
Note that a smaller loss means larger rewards.
The results show that our models perform better in almost all the scenarios as compared to the ES strategy. 
Though the NN models may have less rewards in individual cases, 
in total the NN models earn larger rewards. 
For example, in the smaller budget scenario for service upon arrival, 
our model has a greater loss for bandwidth allocations in classes 1 and 2 
as compared to ES, however, it achieves a smaller loss in total. 
As formulated, our models learn request amount distributions of each class, and then allocate resources optimally to each class such that the budgets are used efficiently.

\renewcommand{\arraystretch}{1.2}
\begin{table} [ht]
\centering
\caption{Simulated data: Loss (-reward) and winners}
\scriptsize{    
\begin{tabular}{lcccccc}
    \hline
          & \multicolumn{6}{c}{Service upon arrival} \\ \cline{2-7}
          &           &  Class    & C1  & C2 & C3   & Total \\ \hline
\multirow{4}{3em}{Smaller budget} & \multirow{2}{1em}{BW} & NN  &4.093E+02&4.935E+02&\textbf{1.510E+03}&\textbf{2.413E+03} \\ 
                                  &                       & ES  &\textbf{2.250E+02}&\textbf{2.250E+02}&8.778E+04&8.823E+04 \\ \cline{2-7}
                                  & \multirow{2}{1em}{VM} & NN  &1.958E+03&\textbf{4.630E+03}&\textbf{5.358E+03}&\textbf{1.195E+04} \\ 
                                  &                       & ES  &\textbf{1.100E+03}&7.620E+04&2.011E+05&2.784E+05 \\ \hline
\multirow{4}{3em}{Small   budget} & \multirow{2}{1em}{BW} & NN  &\textbf{1.428E+02}&\textbf{1.727E+02}&\textbf{4.642E+02}&\textbf{7.796E+02} \\ 
                                  &                       & ES  &2.587E+02&2.587E+02&7.098E+04&7.150E+04 \\ \cline{2-7}
                                  & \multirow{2}{1em}{VM} & NN  &\textbf{6.169E+02}&\textbf{1.423E+03}&\textbf{1.746E+03}&\textbf{3.786E+03} \\ 
                                  &                       & ES  &1.254E+03&4.120E+03&1.244E+05&1.298E+05 \\ \hline
\multirow{4}{3em}{Large budget}   & \multirow{2}{1em}{BW} & NN  &\textbf{1.641E+02}&\textbf{2.107E+02}&\textbf{5.023E+02}&\textbf{8.771E+02} \\ 
                                  &                       & ES  &2.924E+02&2.924E+02&5.418E+04&5.477E+04 \\ \cline{2-7}
                                  & \multirow{2}{1em}{VM} & NN  &\textbf{6.691E+02}&1.536E+03&\textbf{2.018E+03}&\textbf{4.224E+03} \\ 
                                  &                       & ES  &1.408E+03&\textbf{1.437E+03}&4.819E+04&5.103E+04 \\ \hline
\multirow{4}{3em}{Larger budget}  & \multirow{2}{1em}{BW} & NN  &\textbf{1.775E+02}&\textbf{2.328E+02}&\textbf{5.679E+02}&\textbf{9.781E+02} \\ 
                                  &                       & ES  &3.260E+02&3.260E+02&3.739E+04&3.804E+04 \\ \cline{2-7}
                                  & \multirow{2}{1em}{VM} & NN  &\textbf{9.119E+02}&1.594E+03&\textbf{2.180E+03}&\textbf{4.686E+03} \\ 
                                  &                       & ES  &1.562E+03&\textbf{1.562E+03}&2.841E+03&5.965E+03 \\ \hline
    \end{tabular}
    \begin{tabular}{lcccccc}
    \hline
          & \multicolumn{6}{c}{Batch service} \\ \cline{2-7}
          &           &  Class    & C1  & C2 & C3   & Total \\ \hline
\multirow{4}{3em}{Smaller budget} & \multirow{2}{1em}{BW} & NN  &\textbf{6.886E+02}&\textbf{8.217E+02}&\textbf{2.200E+03}&\textbf{3.710E+03}  \\ 
                                  &                       & ES  &1.125E+03&1.126E+03&2.419E+04&2.644E+04  \\ \cline{2-7}
                                  & \multirow{2}{1em}{VM} & NN  &\textbf{2.993E+03}&\textbf{6.880E+03}&\textbf{8.277E+03}&\textbf{1.815E+04}  \\ 
                                  &                       & ES  &5.500E+03&7.707E+03&3.143E+04&4.464E+04  \\ \hline
\multirow{4}{3em}{Small   budget} & \multirow{2}{1em}{BW} & NN  &\textbf{7.371E+02}&\textbf{8.842E+02}&\textbf{2.357E+03}&\textbf{3.978E+03}  \\ 
                                  &                       & ES  &1.293E+03&1.294E+03&1.620E+04&1.879E+04  \\ \cline{2-7}
                                  & \multirow{2}{1em}{VM} & NN  &\textbf{3.179E+03}&7.352E+03&\textbf{8.805E+03}&\textbf{1.934E+04}  \\ 
                                  &                       & ES  &6.270E+03&\textbf{6.878E+03}&1.166E+04&2.481E+04  \\ \hline
\multirow{4}{3em}{Large budget}   & \multirow{2}{1em}{BW} & NN  &\textbf{8.529E+02}&\textbf{9.858E+02}&\textbf{2.577E+03}&\textbf{4.416E+03}  \\ 
                                  &                       & ES  &1.462E+03&1.462E+03&8.587E+03&1.151E+04  \\ \cline{2-7}
                                  & \multirow{2}{1em}{VM} & NN  &\textbf{3.474E+03}&8.125E+03&9.692E+03&\textbf{2.129E+04}  \\ 
                                  &                       & ES  &7.040E+03&\textbf{7.262E+03}&\textbf{8.489E+03}&2.279E+04  \\ \hline
\multirow{4}{3em}{Larger budget}  & \multirow{2}{1em}{BW} & NN  &\textbf{9.089E+02}&\textbf{1.107E+03}&\textbf{2.882E+03}&\textbf{4.898E+03}  \\ 
                                  &                       & ES  &1.630E+03&1.630E+03&3.575E+03&6.836E+03   \\ \cline{2-7}
                                  & \multirow{2}{1em}{VM} & NN  &\textbf{3.870E+03}&8.970E+03&1.064E+04&\textbf{2.348E+04} \\ 
                                  &                       & ES  &7.809E+03&\textbf{7.892E+03}&\textbf{8.389E+03}&2.409E+04  \\ \hline
    \end{tabular}
    }
\label{table:sim-loss}
\end{table}

\subsection{Real Data}
The models are further validated in real settings by conducting trace driven experiments using a Facebook cluster compute workload trace for CPU and a 4G LTE trace for bandwidth (details in Section~\ref{section:data}). 
Assuming the models control both bandwidth and computing requests concurrently, 
the traces are combined as described in Section \ref{section:data}. 
For the bandwidth trace, we used three classes: Bus, Car, and Train from the data. For the compute trace, we use the traces labeled FB-2009-0, FB-2009-1, and FB-2010-0 for each class separately. 
For bandwidth requests, it is assumed that the maximum size of the request amounts since the last arrival should be serviced. 
Also, given the limited number of traces, we assume that the requests are recurrent.
To prevent divergence during training, we appropriately scaled down the original values for compute and bandwidth in the traces. 
The traces are split into two sets: 90 \% of the trace is used to train the model, and 10 \% is used for testing. 
The experimental setup is similar to the simulated data experiments including the settings for the following parameters: hyper-parameters,
weights for delays and resource usage costs, service time interval for batch service. 
Each neural agent has three hidden layers with 1,000 units. Were we also use the Adam optimizer with learning rate 0.001.

\begin{figure*} 
    \centering
  \subfloat[Resource allocation: Class 1\label{1a}]{%
       \includegraphics[width=0.45\linewidth]{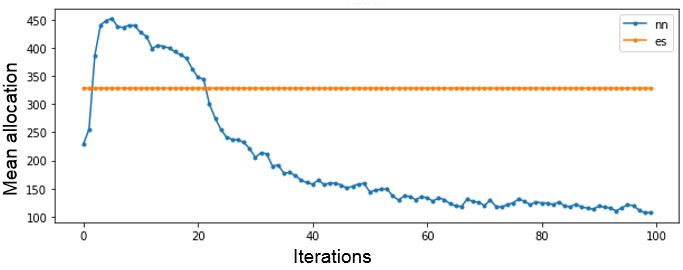}}
  \subfloat[Buffer level: Class 1\label{1b}]{%
        \includegraphics[width=0.45\linewidth]{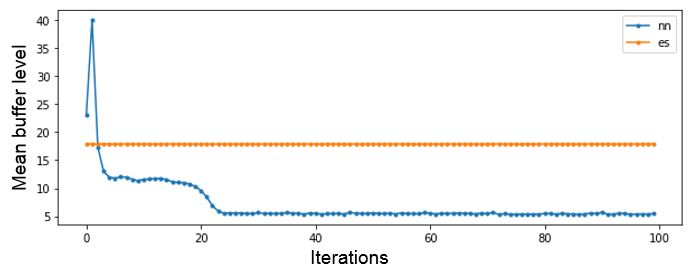}}
        \hfill
  \subfloat[Resource allocation: Class 2\label{2a}]{%
       \includegraphics[width=0.45\linewidth]{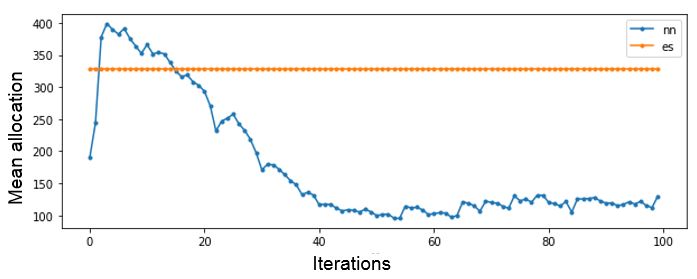}}
  \subfloat[Buffer level: Class 2\label{2b}]{%
        \includegraphics[width=0.45\linewidth]{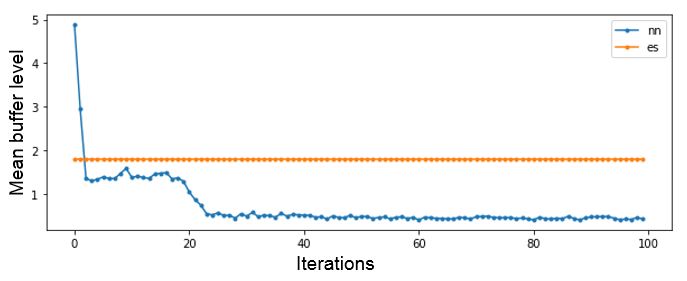}}
        \hfill
  \subfloat[Resource allocation: Class 3\label{3a}]{%
       \includegraphics[width=0.45\linewidth]{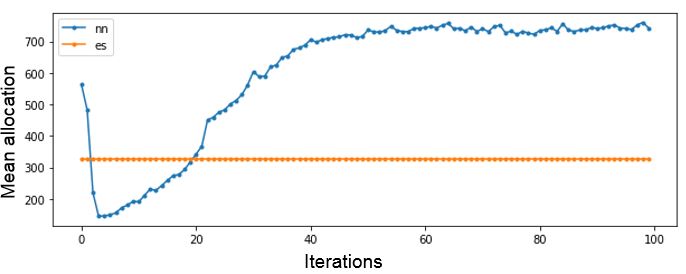}}
  \subfloat[Buffer level: Class 3\label{3b}]{%
        \includegraphics[width=0.45\linewidth]{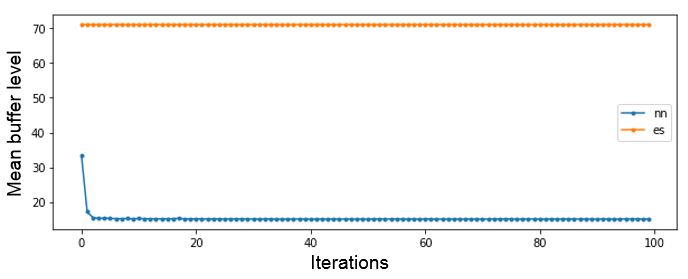}}
  \caption{Real data: Mean resource allocations, and buffer levels of three classes (large budget, service upon arrival, VM)}
  \label{real:fig} 
\end{figure*}

Table \ref{table:real-loss} shows the results using the real data trace driven experiments, of the proposed 
models (\textbf{NN}) compared with the equal slicing strategy (\textbf{ES}),
in terms of the expected rewards for each class and resource. We denote the winners in \textbf{bold}.
%with the winners shown in \textbf{bold}. 
As seen with the simulated data, the results show that our models perform better in almost all the scenarios as compared to the ES strategy. 
Though the NN models may have less rewards in individual cases, 
in total the NN models earn larger rewards.
For example, in the smaller budget scenario for service upon arrival our model has a greater loss for VM allocations in class 1 and 3 as compared to the ES startegy, however,
it achieves a smaller total loss as it saves a lot in class 3. 
This can be explained by looking at the training phase shown in Figure \ref{real:fig}, 
in that our models learn the request amount distributions of all classes and 
then allocate resources accordingly. 
During training, our models allocate resources differently to each class, and buffer levels also decrease.

Based on experimental results on both simulated and real data, we can conclude that our models can learn efficient resource allocation policies for the coupled dynamic resource allocation problem for network slicing and outperform the baseline strategy.

%Based on experimental results on both simulated and real data, we can conclude that our models successfully solve the dynamic multiple-resource allocation problem.

\begin{table} [ht]
\centering
\caption{Real data: Loss (-reward) and winners}
\scriptsize{    
\begin{tabular}{lcccccc}
    \hline
          & \multicolumn{6}{c}{Service upon arrival} \\ \cline{2-7}
          &           &  Class    & C1  & C2 & C3   & Total \\ \hline
\multirow{4}{3em}{Smaller budget} & \multirow{2}{1em}{BW} & NN  &1.170E+05&\textbf{2.933E+05}&\textbf{4.558E+04}&4.559E+05  \\ 
                                  &                       & ES  &\textbf{1.140E+05}&2.935E+05&4.847E+04&\textbf{4.559E+05}  \\ \cline{2-7}
                                  & \multirow{2}{1em}{VM} & NN  &\textbf{4.679E+02}&6.658E+02&\textbf{6.776E+03}&\textbf{7.909E+03}  \\ 
                                  &                       & ES  &9.514E+02&\textbf{6.913E+01}&1.367E+06&1.368E+06  \\ \hline
\multirow{4}{3em}{Small   budget} & \multirow{2}{1em}{BW} & NN  &4.585E+02&\textbf{1.878E+03}&6.056E+02&\textbf{2.942E+03}  \\ 
                                  &                       & ES  &\textbf{2.387E+01}&1.318E+05&\textbf{1.793E+01}&1.319E+05  \\ \cline{2-7}
                                  & \multirow{2}{1em}{VM} & NN  &\textbf{9.702E+01}&\textbf{9.266E+01}&\textbf{5.609E+02}&\textbf{7.506E+02}  \\ 
                                  &                       & ES  &1.718E+02&1.719E+02&9.897E+02&1.333E+03  \\ \hline
\multirow{4}{3em}{Large budget}   & \multirow{2}{1em}{BW} & NN  &9.017E+00&\textbf{1.379E+01}&\textbf{7.341E+00}&\textbf{3.015E+01}  \\ 
                                  &                       & ES  &\textbf{7.546E+00}&1.418E+03&7.546E+00&1.433E+03  \\ \cline{2-7}
                                  & \multirow{2}{1em}{VM} & NN  &\textbf{9.510E+01}&\textbf{1.211E+02}&8.374E+02&\textbf{1.054E+03}  \\ 
                                  &                       & ES  &3.231E+02&3.232E+02&\textbf{7.204E+02}&1.367E+03  \\ \hline
\multirow{4}{3em}{Larger budget}  & \multirow{2}{1em}{BW} & NN  &\textbf{7.521E+00}&\textbf{1.321E+01}&\textbf{7.910E+00}&\textbf{2.864E+01}  \\ 
                                  &                       & ES  &9.548E+00&4.866E+01&9.548E+00&6.775E+01  \\ \cline{2-7}
                                  & \multirow{2}{1em}{VM} & NN  &\textbf{8.966E+01}&\textbf{1.095E+02}&1.271E+03&\textbf{1.470E+03}  \\ 
                                  &                       & ES  &4.757E+02&4.757E+02&\textbf{7.341E+02}&1.685E+03  \\ \hline
    \end{tabular}
    \begin{tabular}{lcccccc}
    \hline
          & \multicolumn{6}{c}{Batch service} \\ \cline{2-7}
          &           &  Class    & C1  & C2 & C3   & Total \\ \hline
\multirow{4}{3em}{Smaller budget} & \multirow{2}{1em}{BW} & NN  &1.663E+04&\textbf{2.795E+04}&1.312E+04&\textbf{5.771E+04}  \\ 
                                  &                       & ES  &\textbf{1.584E+04}&3.176E+04&\textbf{1.129E+04}&5.888E+04  \\ \cline{2-7}
                                  & \multirow{2}{1em}{VM} & NN  &2.604E+03&9.454E+03&\textbf{8.106E+04}&\textbf{9.311E+04} \\ 
                                  &                       & ES  &\textbf{1.246E+02}&\textbf{1.361E+02}&1.553E+05&1.556E+05  \\ \hline
\multirow{4}{3em}{Small   budget} & \multirow{2}{1em}{BW} & NN  &1.283E+04&\textbf{1.594E+04}&9.770E+03&\textbf{3.854E+04 } \\ 
                                  &                       & ES  &\textbf{1.108E+04}&2.081E+04&\textbf{6.741E+03}&3.862E+04  \\ \cline{2-7}
                                  & \multirow{2}{1em}{VM} & NN  &\textbf{8.109E+02}&\textbf{6.816E+02}&2.009E+03&\textbf{3.501E+03 } \\ 
                                  &                       & ES  &1.163E+03&1.163E+03&\textbf{1.342E+03}&3.669E+03  \\ \hline
\multirow{4}{3em}{Large budget}   & \multirow{2}{1em}{BW} & NN  &7.517E+03&\textbf{1.074E+04}&6.762E+03&\textbf{2.502E+04}  \\ 
                                  &                       & ES  &\textbf{6.588E+03}&1.620E+04&\textbf{2.260E+03}&2.505E+04  \\ \cline{2-7}
                                  & \multirow{2}{1em}{VM} & NN  &\textbf{1.549E+03}&\textbf{9.145E+02}&4.148E+03&\textbf{6.612E+03} \\ 
                                  &                       & ES  &2.204E+03&2.204E+03&\textbf{2.234E+03}&6.642E+03  \\ \hline
\multirow{4}{3em}{Larger budget}  & \multirow{2}{1em}{BW} & NN  &3.229E+03&\textbf{5.697E+03}&2.654E+03&\textbf{1.158E+04}  \\ 
                                  &                       & ES  &\textbf{2.109E+03}&1.169E+04&\textbf{7.079E+01}&1.387E+04  \\ \cline{2-7}
                                  & \multirow{2}{1em}{VM} & NN  &\textbf{2.670E+03}&\textbf{2.145E+03}&4.919E+03&\textbf{9.734E+03 } \\ 
                                  &                       & ES  &3.245E+03&3.245E+03&\textbf{3.252E+03}&9.741E+03  \\ \hline
    \end{tabular}
    }
    \label{table:real-loss}
\end{table}

\section{Related Work}
\label{section:related-work}
There is considerable literature on communication network resource allocation problems
including network slicing using RL, and more recently, 
deep RL approaches; these are categorized as follows (TABLE \ref{table:related-work}).
\begin{table}[H]
\centering
\caption{Current approaches for resource allocation}
\begin{tabular}{|l|l|l|}
\hline
                                                  & \multicolumn{1}{c|}{Resource Allocation}                                                                                                                                & \multicolumn{1}{c|}{Network Slicing}                                                                                        \\ \hline
RL                                                & \begin{tabular}[c]{@{}l@{}}Bandwidth Allocation\\ \cite{13Elwalid1995}, \cite{14Hetzer2006}, \cite{15Tong2000}, \cite{16Hui2003},  \cite{17Nordstrom1995}\\ Compute Allocation \cite{18Jamshidi2015}, \cite{19Benifa2018}\end{tabular}                                                     & \begin{tabular}[c]{@{}l@{}}Q-learning \cite{2Bega2017}\\ Genetic optimization\\ \cite{3Han2018}\end{tabular} \\ \hline
\begin{tabular}[c]{@{}l@{}}Deep\\ RL\end{tabular} & \begin{tabular}[c]{@{}l@{}}Cognitive radio networks \cite{9He2017} \\ 
Cloud radio access networks \cite{10Xu2017} \\
Vehicular ad hoc networks \cite{11He2017}
\end{tabular} & \begin{tabular}[c]{@{}l@{}}Deep RL for network\\ slicing \cite{4Zhao2018} \end{tabular}                                                   \\ \hline
\end{tabular}
\label{table:related-work}
\end{table}

In \cite{13Elwalid1995} the authors study the the admissibility of
variable bit rate (VBR) traffic and bandwidth allocation in buffered ATM networks. 
%formulating the problem as a
%single-resource statistical-multiplexing problem.
\cite{17Nordstrom1995} presents an adaptive link allocation scheme for ATM networks 
formulated as semi-Markov Decision Problems (SMDPs).
%, that are directly adapted by RL.
In \cite{14Hetzer2006} the authors propose adaptive bandwidth  planning  using  RL for
optimal scheduling  considering  QoS  parameter patterns  as  feedback 
from  the environment. 
%where the  approach  is  based on constrained scheduling
%algorithms.
Adaptive call admission control (CAC) in multimedia networks is addressed in \cite{15Tong2000} via RL, formulating the problem as a constrained SMDP
maximizing revenue and meeting packet-level and call-level QoS constraints. 
The work in \cite{16Hui2003} uses RL for adaptive provisioning of differentiated services networks
for per hop behavior when the bandwidth required is not known at the time of connection admission.
Cloud compute resource allocation is addressed in \cite{18Jamshidi2015} 
where the authors propose learning  adaptation  rules  by a cloud controller that learns and modifies fuzzy  rules  at  run-time for scaling cloud 
resources.
\cite{19Benifa2018} addresses the same issue,  
basing their work on the RL-SARSA algorithm that learns the environment
dynamically in parallel and allocates the resources.

Network resource allocation using deep RL models have been considered in various domains such as cognitive radio for smart cities~\cite{9He2017}, cloud RAN~\cite{10Xu2017}, and VANET~\cite{11He2017}.
%is addressed in the following papers:
%cognitive radio networks for smart cities \cite{9He2017}, 
%cloud radio access networks \cite{10Xu2017}, and
%vehicular ad hoc networks \cite{11He2017}.
In \cite{9He2017} the authors propose an integrated framework for dynamic 
orchestration of networking, caching, and computing resources for smart city applications, where the
algorithm performance is evaluated through simulations (real data is not used).
In \cite{2Bega2017} and \cite{3Han2018}, the authors propose  algorithms for slicing admission strategy optimization in 5G networks using Q-learning and genetic algorithms; however, they consider a binary decision
mechanism where declined requests are dropped.
In \cite{4Zhao2018}, which is closest to our work, 
the authors formulate the network slicing problem with deep RL
for two typical resource management scenarios: radio
resource slicing for a base station with multiple services; and
%the reward is defined as a weighted sum of spectrum efficiency and QoE.
priority-based core network slicing, 
with multiple service function chains requiring different compute
resources and waiting time.
%the reward is defined as the sum of waiting time in different SFCs.
However, our work is different from \cite{4Zhao2018} in several critical aspects ---
our models address the
issue of allocating multiple resources simultaneously, 
solve constrained problems by introducing buffers, and 
consider both service upon arrival and batch service.
Furthermore, the performance of our approach is validated with simulated and real data.

\section{Conclusions}
\label{section:conclusions}
In this paper we proposed a new deep RL framework for network slicing with heterogeneous resource requirements and finite capacity which can deal with highly dynamic traffic demands from network users.
Experiments using both synthetic and real workload driven traces show that our system performs well compared to a baseline equal-slicing strategy. 
Our RL algorithm can be trained offline using the simulated and trace data, and the learned policies can be used in real-time for end-to-end 5G slicing systems.
Taking a more holistic view of network slicing, we plan to extend this work
to lifecycle management of network slices to include
creation, activation and deactivation, and elasticity for slices.
We plan to improve our RL architecture by exploring different learning algorithms such as Actor-Critic methods~\cite{Konda:2002:AA:936987}, and DQN~\cite{mnih-atari-2013}. We also plan to deploy the algorithms in a real 5G network slicing test-bed.

\bibliographystyle{ieeetr}
\bibliography{references}
\end{document}